\documentclass[prb,aps,twocolumn,superscriptaddress]{revtex4-2}
\usepackage{graphicx}
\usepackage{dcolumn}
\usepackage{bm}
\usepackage{lipsum}
\usepackage{ragged2e}
\usepackage{braket}
\usepackage{bbm}
\usepackage{lipsum}
\usepackage{color}
\usepackage{xcolor}
\usepackage{amsfonts}
\usepackage{amsmath}
\usepackage{amsmath,amssymb,latexsym}
\usepackage{wasysym,stmaryrd,marvosym}
\usepackage{mathrsfs}
\usepackage{dsfont}
\usepackage{float}
\providecommand{\keywords}[1]{\textbf{\textit{Index terms---}} #1}
\usepackage[mathlines]{lineno}
\usepackage[colorinlistoftodos]{todonotes}
\usepackage[colorlinks=true, allcolors=blue]{hyperref}
\newcommand{\RC}[1]{{\color{violet}{#1}}} %changes by Ramon

\begin{document}
%\setcitestyle{super}
\newcommand{\cnyn}{Centro de Nanociencias y Nanotecnolog\'ia,
Universidad Nacional Aut\'onoma de M\'exico, Apartado Postal 2681, 22800, Ensenada, Baja California, M\'exico.}
\newcommand{\uabc}{Facultad de Ciencias, Universidad Aut\'onoma de Baja California, Apartado Postal 1880, 22800 Ensenada, Baja California, M\'exico}
\newcommand{\ohio}{Department of Physics and Astronomy, Ohio University, Athens, Ohio 45701, USA}

\title{Hyperbolic plasmons in massive tilted 2D Dirac materials}
%\title{Hyperbolic plasmons in gapped 2D Dirac fermions with tilt}

\author{M. A. Mojarro}
\affiliation{\ohio}\affiliation{\uabc}
\author{R. Carrillo-Bastos}
\affiliation{\uabc}
\author{Jes\'us A. Maytorena}
\email{jesusm@ens.cnyn.unam.mx}
\affiliation{\cnyn}

\date{\today}

\begin{abstract}
%\RC{I changed to active voice}
We explore topological transitions in the type of propagation of surface electromagnetic modes in massive anisotropic tilted Dirac systems. The presence of tilting and mass gives rise to an indirect band gap that strongly modifies the joint density of states compared to the gapless system. New van Hove singularities appear, and the interplay between intra and interband transitions leads to an anisotropic optical conductivity with imaginary parts acquiring opposite signs in orthogonal directions, opening the possibility of having hyperbolic propagation of plasmons. Isofrequency contours and low plasmon losses, as obtained from the dispersion relation, show that transitions between purely anisotropic quasi-elliptical and well-defined, highly directional, hyperbolic modes are attainable only when tilt and mass coexist. Fine-tuning in the THz regime suggests that this might be useful as a dynamical probe of concurring tilt and mass in Dirac materials.

\end{abstract}


\maketitle
The recent emergence of natural two-dimensional hyperbolic materials\cite{TLow_NatureMatt2017,TLowCh5} offers a way to circumvent the fabrication challenges of hyperbolic metasurfaces\cite{AluPRL2015} and the constrictions on dispersion relation that brings a sizeable structural periodicity\cite{TLow2016}. Moreover, these 2D materials have shown a hyperbolic plasmonic behavior in a wide range of frequencies from mid-infrared to UV, accompanied by high tuneability with doping and gating\cite{sun2014indefinite}. For example, experiments on thin films of WTe$_2$ demonstrated a hyperbolic frequency range of 13-18 THz\cite{wang2020van} originated from in-plane intraband anisotropic transitions\cite{Torbatian-PRApp.14.044014}; and first-principles calculations showed a second window around 241 THz, associated with resonant anisotropic interband transitions via band-nesting\cite{TLow2020PRBr}. In contrast, MoTe$_2$, another transition-metal telluride, is a natural hyperbolic material with low losses across the visible and the ultraviolet region\cite{Edalati-PRMat.4.085202} (725-1450 THz). Still, the hyperbolicity condition now relies upon the in-plane and out-of-plane components of the dielectric tensor. Similar mechanisms are behind the indefinite behavior of electride materials\cite{guan2017tunable} and Van der Waals crystals\cite{zheng2019mid} that present hyperbolic windows in the infrared, and layered hexagonal crystals\cite{ebrahimian2021natural} that are hyperbolic in the visible and UV range. Other examples of two-dimensional hyperbolic materials include Black phosphorous with two hyperbolic windows\cite{correas2016black,van2019tuning} in the infrared (80 THz) and visible range (677 THz), carbon phosphide\cite{dehdast2021tunable} that is hyperbolic in the infrared, and  8-$Pmmn$ borophene\cite{Torbatian-PRB.104.075432}, MoOCl$_2$\cite{zhao2020highly} and nodal-line semimetals\cite{nodal2022,nodal2-PRL.119.187401}. 
In all these systems, hyperbolicity results from the intrinsic anisotropic arrangement of atoms in the material and its corresponding anisotropic interplay of intra and interband optical responses. Therefore, typical isotropic Dirac materials like graphene do not display hyperbolic plasmons; even with anisotropic Fermi velocity\cite{DasSarma2021} or tilting\cite{verma2017effect}, the low-losses plasmons dispersion for these materials is still elliptical. However, in a recent study, we show that massive tilted Dirac systems, like the metal-organic $\alpha$-(BEDT-TTF)$_{2}$I$_{3}$\cite{uykur2019optical,Osada-experimental},
present a directional optical response due to the combination of tilting and gap\cite{MRJ_1}, thus, opening the possibility of hosting low-loss hyperbolic plasmons for Dirac materials. Consequently, in this manuscript, we report a new scenario of hyperbolic propagation in 2D Dirac systems involving tilt {\it and} mass (in the tens of THz), and show that if one of these elements is absent, the plasmon propagation is of purely anisotropic elliptic type strictly, without any transition to a hyperbolic regime.

We consider a 2D gapped anisotropic Dirac system with low-energy Hamiltonian \cite{MRJ_1,Stegmann-PRB.105.045401,Gosh2020,Rostami-Hall2020} 
\begin{equation} \label{H1}
H_{\xi}({\bf k})=\xi(\hbar v_tk_y\mathds{1}+\hbar v_xk_x\sigma_x+\xi\hbar v_yk_y\sigma_y)+\Delta\sigma_z,
\end{equation}
where $\sigma_i$  are  Pauli matrices defined on the pseudospin space, the carrier velocities $v_x, v_y$ account for the
anisotropy of the model, and ${\bf k}=k_x{\bf\hat{x}}+k_y{\bf\hat{y}}$ is the electron wave vector for states in the vicinity of the 
$K\,(K')$ point with valley index $\xi=+\,(-)$; $\mathds{1}$ is the  $2\times 2$ identity matrix. 
The system includes a mass $\Delta>0$ in each valley and an amount of tilting $\xi v_t$ along
the $k_y$-axis. 
% {original} The system includes a mass $\Delta>0$ in each valley and an amount of tilting $\xi v_t$ of the energy bands along the $k_y$-axis, determined by the velocity $v_t$.
In the absence of mass, the model describes some 2D graphene-like materials
and organic conductors \cite{goerbig2008tilted,MagnetoplasmonsPRB,rostamzadeh2019large,verma2017effect}. Taking $v_x=v_y=v_F$ reduces Eq. \eqref{H1} to graphene Hamiltonian, 
while $8$-$Pmmn$ borophene Hamiltonian is recovered  with the values\cite{zabolotskiy2016strain} $v_x=0.86\,v_F,v_y=0.69\,v_F$, and $v_t=0.32\,v_F$ with $v_F=10^6\,$m/s. 
A realistic example of the complete model is the organic conductor $\alpha$-(BEDT-TTF)$_2$I$_3$, a well recognized 2D massive Dirac fermion system
with a pair of tilted Dirac cones under hydrostatic pressure close to a critical pressure\cite{Osada-experimental,TheoryAndExp,ExpAnis,OsadaCurrentInduced}.

\begin{figure}
    \centering
    \includegraphics[scale=0.28]{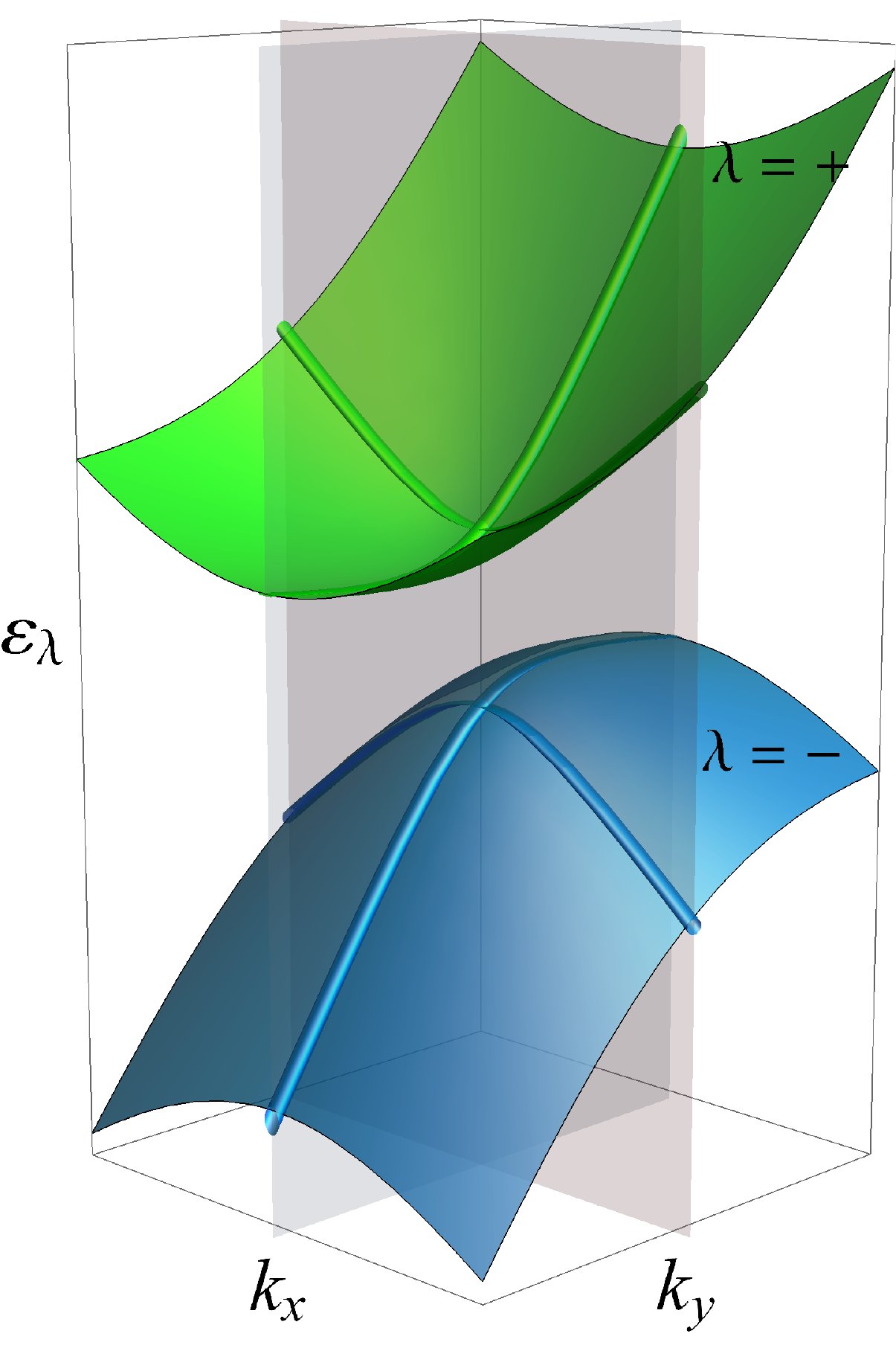}
    \includegraphics[scale=0.28]{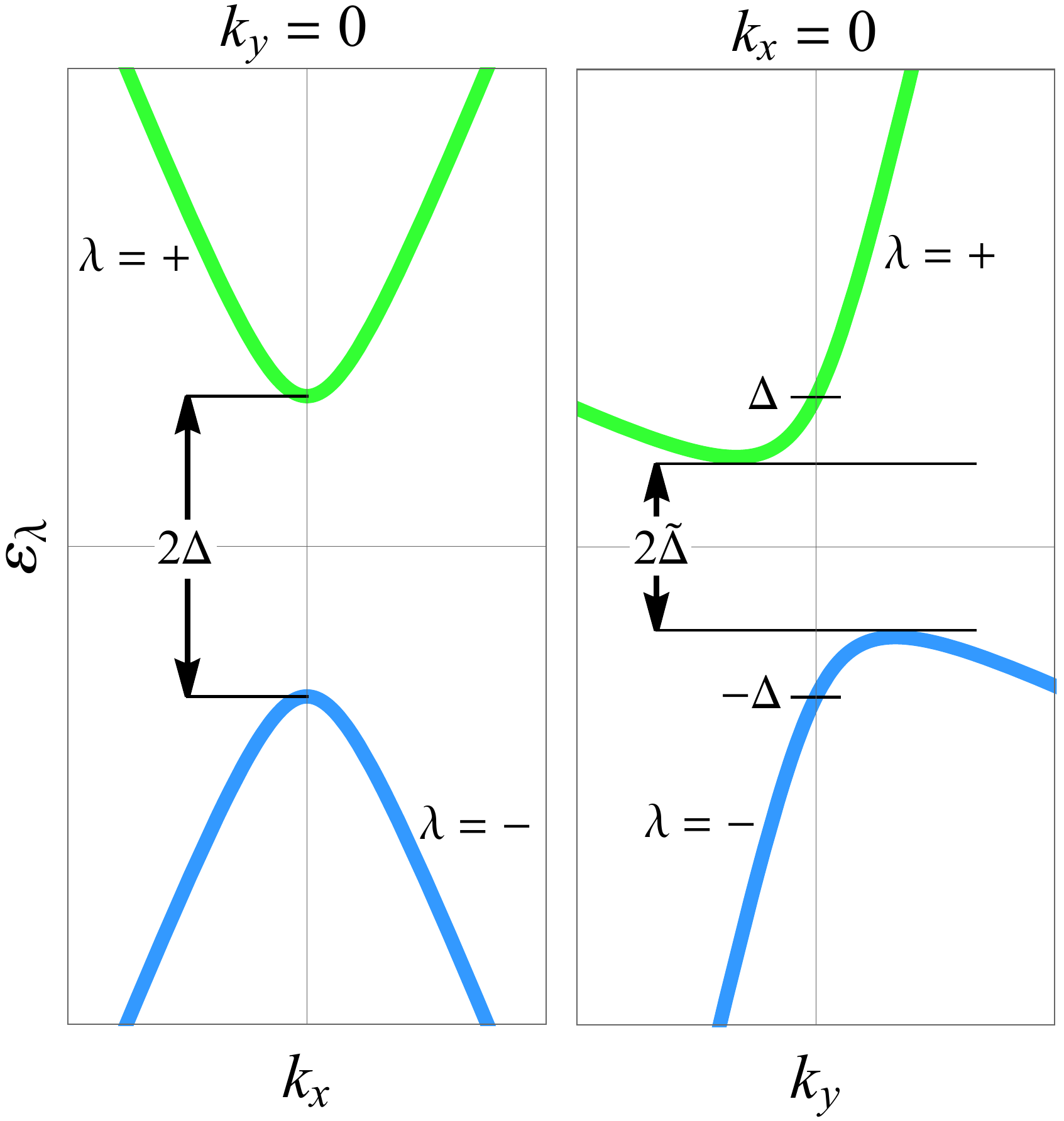}
    \caption{Energy bands at a $\xi=+$ valley of a massive tilted
    Dirac system, and two cuts with planes $k_y=0$ and $k_x=0$. Along the $k_x$-axis the bands are separated
    by $2\Delta$, but along the $k_y$-axis the system presents the minimum separation $2\tilde{\Delta}$ ($<2\Delta$) of indirect nature. This represents an additional source of anisotropy and modifies the interplay between intra and interband contributions to the optical response.
    }
    \label{fig:bands}
\end{figure}

The energy-momentum dispersion reads
\begin{equation} \label{bands}
\varepsilon_{\xi,\lambda}(k_x,k_y)=\xi\alpha_tk_y+\lambda\sqrt{\alpha^2_xk_x^2+\alpha_y^2k_y^2+\Delta^2},
\end{equation}
where $\,\alpha_i=\hbar v_i\,(i=x,y,t,F)$ and the index $\lambda=\pm$ labels the conduction ($\lambda=+$) and the valence  ($\lambda=-$) bands in each valley $\xi=\pm$.
A notable feature of the model is that the simultaneous presence of tilt and mass produces an indirect gap in each valley
around the nominal Dirac point (Fig. \ref{fig:bands}). 
Indeed, the conduction branches $\varepsilon_{\xi,+}({\bf k})$ present a minimum at ${\bf k}=-\xi Q{\bf\hat{y}}$,
while the valence branches $\varepsilon_{\xi,-}({\bf k})$ display a maximum at ${\bf k}=+\xi Q{\bf\hat{y}}$, 
where $\alpha_yQ=\Delta(\gamma/\sqrt{1-\gamma^2})$ with the tilting parameter $\gamma=v_t/v_y\,(0\leqslant \gamma <1)$.
Thus, a Fermi level within the gap reads $\varepsilon_F<\tilde{\Delta}$, where $\tilde{\Delta}=\Delta\sqrt{1-\gamma^2} < \Delta$.
This means that a new scenario is possible with the Fermi level lying in the ``indirect zone''
$\tilde{\Delta}<\varepsilon_F<\Delta$. This has a striking effect on the spectrum of interband transitions\cite{MRJ_1}.
(1) For the Fermi level
within the gap, the joint density of states (JDOS) displays a graphene-like linear dependence on the exciting energy $\hbar\omega$ above the threshold $2\Delta$,
but involving the geometric mean  velocity $\sqrt{v_xv_y}$ instead of Fermi velocity $v_F$.
%\cite{MRJ_1} 
%[referencia, PRB].  
(2) On the other hand, for
$\varepsilon_F>\Delta$, the absence of particle-hole symmetry leads to the appearance of two critical energies 
\begin{equation}
\hbar\omega_{\pm}=2\left[\varepsilon_F\pm\gamma\sqrt{\varepsilon^2_F-\tilde{\Delta}^2}\right]/(1-\gamma^2),
\end{equation}
instead  of
the unique absorption edge $2\varepsilon_F$\RC{,} characteristic of monolayer graphene. Below $\omega_-$ the JDOS vanishes, and above $\omega_+$ it presents the typical linear behavior.
Between $\omega_-$ y $\omega_+$ the behavior is no longer lineal and,  globally, the JDOS has a behavior which resembles that of the ungapped 
8-$Pmmn$ borophene\cite{verma2017effect}.
(3) In contrast, when Fermi level lies in the indirect zone, the JDOS displays a set of three van Hove singularities (at $2\Delta$ and $\hbar\omega_{\pm}$)
and a significant overall size reduction, due to a drastic shrinking of the momentum space available for direct transitions, caused by the indirect nature of the gap. In the same way, the intraband spectral weight is also anisotropic with a nonlinear dependence on the Fermi energy\cite{MRJ_1}.

\begin{figure}
    \centering
    \includegraphics[scale=0.45]{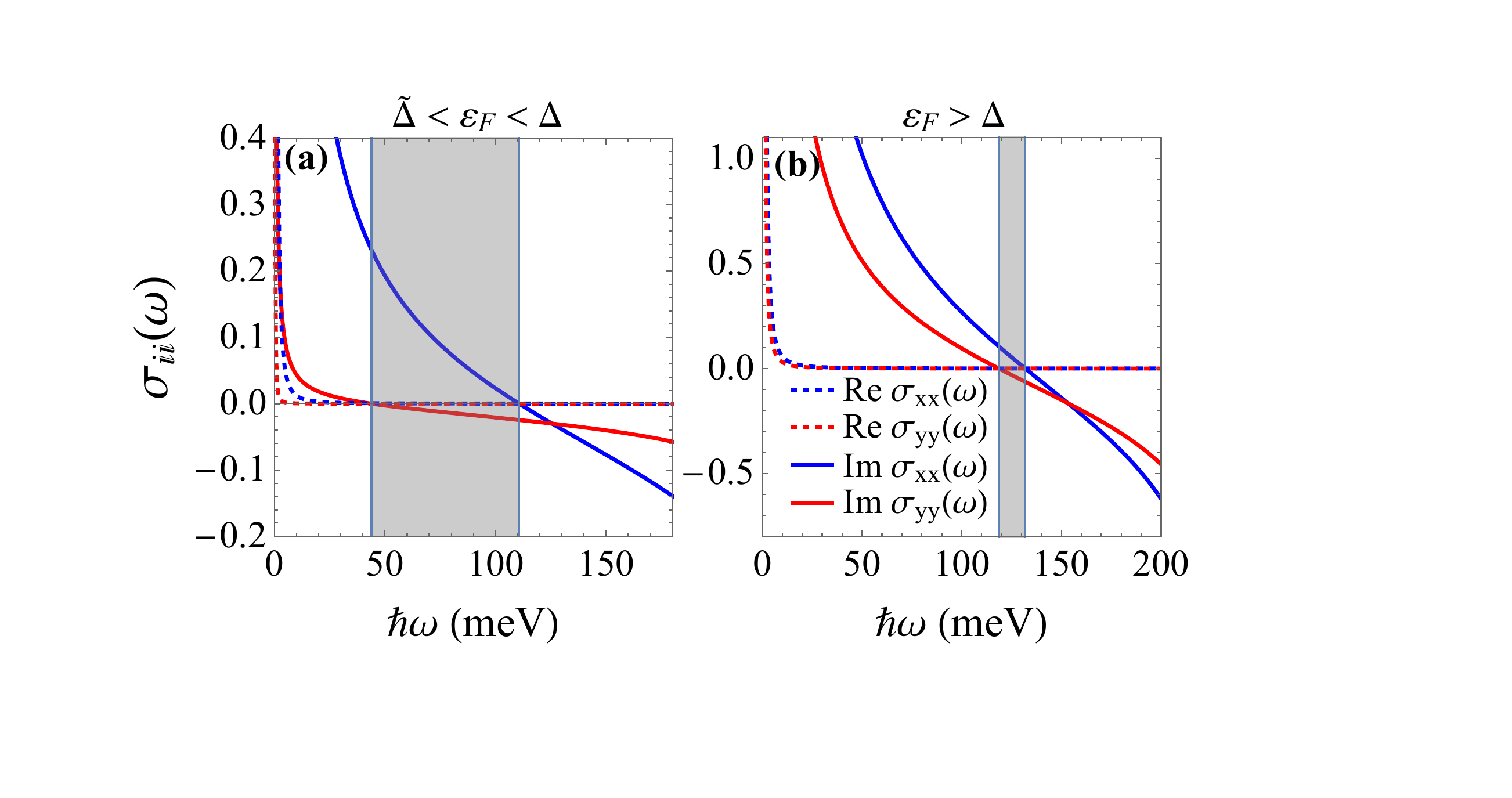}
    \caption{Optical conductivities $\sigma_{xx}(\omega)$ and $\sigma_{yy}(\omega)$ (in units of $2e^2/h$), 
    below the onset of interband transitions.
    (a) $\tilde{\Delta}<\varepsilon_F<\Delta$, 
    (b) $\varepsilon_F>\Delta$. We take
    $\Delta=110$ meV. The shaded areas indicate hyperbolic regimes. 
    %$($\varepsilon_F=103$)($\varepsilon_F=120$)\hbar\omega_-=223$ meV.
    }
    \label{fig2}
\end{figure}

These spectral characteristics reveal that the intraband and interband contributions to the optical conductivity tensor $\sigma_{ij}(\omega)$
can be strongly modified by locating the Fermi level properly. This offers the opportunity to control the form of propagation of surface electromagnetic modes. 
In particular, as we will see below,  the conditions for hyperbolic plasmons become accessible.
As is well known,\cite{TLow_NatureMatt2017,AluPRL2015,Alu} a 2D material can support hyperbolic dispersion relation of plasmons when $\text{Im}[\sigma_{xx}(\omega)]\text{Im}[\sigma_{yy}(\omega)]<0$. 
The combination of time-reversal symmetry and broken inversion symmetry of the model (\ref{H1}) 
leads to an anisotropic response\cite{MRJ_1} 
$\sigma_{ij}(\omega)=\delta_{ij}[\sigma_{xx}(\omega)\delta_{ix}+\sigma_{yy}(\omega)\delta_{iy}]$. 
Fig. \ref{fig2}(a) shows the $xx$ and $yy$ components when the Fermi level lies in the indirect zone $\tilde{\Delta}<\varepsilon_F<\Delta$, 
as calculated from Kubo formula. A well defined hyperbolic region of frequencies exists 
where the losses are negligible, below the onset $2\Delta$ of interband transitions and above the Drude peak. 
A hyperbolic region also appears for $\varepsilon_F>\Delta$, as can be seen in Fig. \ref{fig2}(b), its width narrowing as $\Delta/\varepsilon_F$ approaches unity. 

\begin{figure}
    \includegraphics[scale=0.58]{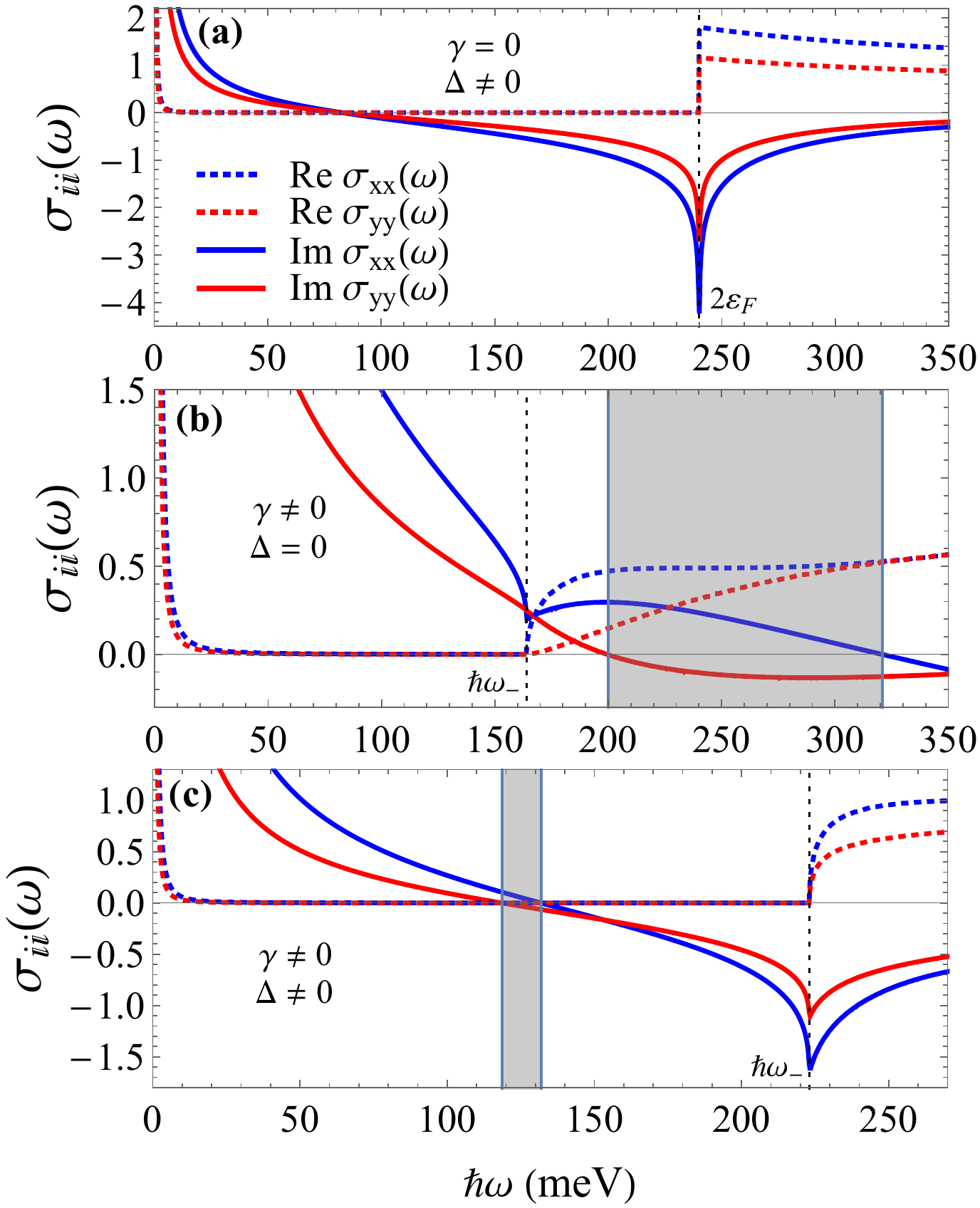}
    \caption{Optical conductivities $\sigma_{xx}$ and $\sigma_{yy}$ (in units of $2e^2/h$), for Dirac cones with (a) gap, (b) tilt, and (c) gap and tilt. The shaded zones indicate the hyperbolic regimes. For (a) and (c) we take $\Delta=110$ meV and $\varepsilon_F>\Delta$. 
    The vertical dashed line denote the onset $\hbar\omega_-$ for interband transitions in each case; it is blue shifted when a gap is introduced in the tilted system.
     %$\Delta=110$ meV, $\varepsilon_F=120$ meV.
     }
    \label{fig3}
\end{figure}

It is interesting to note that for a system with $\Delta\geqslant 0$ and $\gamma=0$, the sign change of the imaginary parts of the longitudinal conductivities occurs exactly at the same frequency, and therefore $\text{Im}[\sigma_{xx}(\omega)]\text{Im}[\sigma_{yy}(\omega)]\geqslant 0$ in the whole frequency range (Fig. \ref{fig3}(a)). On the other hand, when only tilting is present, $\Delta=0$ and $\gamma\neq 0$, the condition for hyperbolicity is satisfied but above the onset for interband absorption only (Fig. \ref{fig3}(b)), implying damped hyperbolic modes.

The calculations reveal that only the simultaneous presence of mass and tilting provides a way to asymmetrically change the magnitude of the interband response versus the (positive) intraband contribution to $\text{Im}[\sigma_{ii}]$ in order
to fulfill the hyperbolicity condition in a spectral region of low lossless. For $\varepsilon_F>\Delta$, the opening of a gap in the tilted system cause a blue shift of the absorption edge $\hbar\omega_-$,
increasing the frequency region between the Drude absorption and Landau damping (Fig.\,\ref{fig3}(c)). This is in contrast to the gapless case, where the onset for interband transitions can be arbitrarily small as $\varepsilon_F\to 0$, as in monolayer graphene or
borophene 8-$Pmmn$\cite{verma2017effect}. Similarly, when
$\tilde{\Delta}<\varepsilon_F<\Delta$, a threshold $2\Delta$ appears for single-particle excitations, which also
left spectral space for undamped hyperbolic plasmons.

\begin{figure}
    \centering
    \includegraphics[scale=0.41]{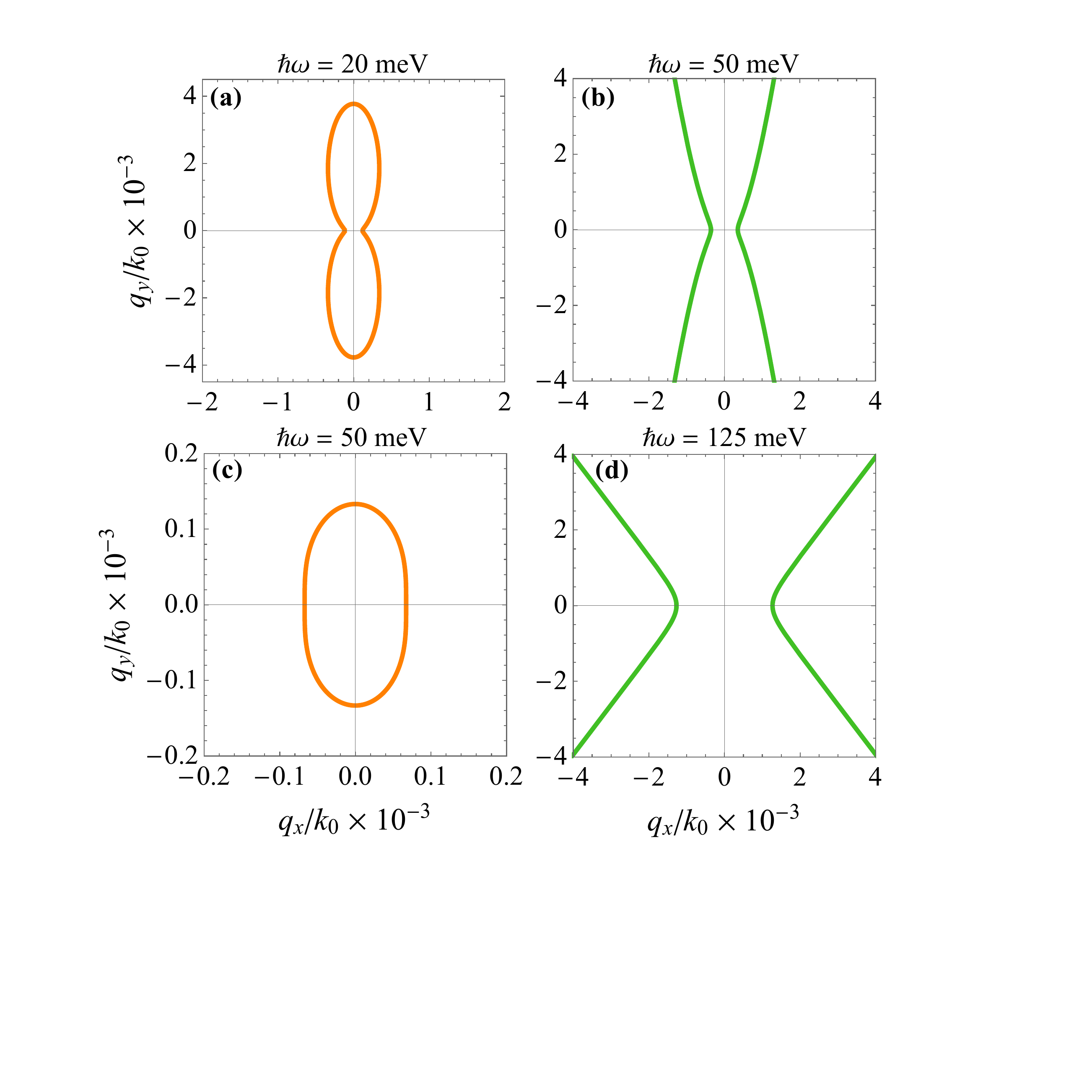}
    \caption{Isofrequency contours $\omega(q_x,q_y)=\text{constant}$ illustrating the plasmon propagation when
    $\tilde{\Delta}<\varepsilon_F<\Delta$ ((a), (b)), and
     $\varepsilon_F>\Delta$ ((c), (d)). The closed (open) curves correspond to a frequency lying outside (inside) the hyperbolic regions displayed in Fig. \ref{fig2}. The group velocity field points perpendicularly to these contours, indicating the direction of energy propagation.}
    \label{fig4}
\end{figure}

The dispersion relation of surface waves localized in the $z$-direction and propagating along the direction of the wave vector ${\bf q}=q_x{\bf\hat{x}}+q_y{\bf\hat{y}}=q(\cos\phi{\bf\hat{x}}+\sin\phi{\bf\hat{y}})$  
on the $xy$ plane is given by\cite{Hanson,Alu,TLowCh5}
$\tilde{\sigma}_{xx}(q_x^2-k_0^2)+\tilde{\sigma}_{yy}(q_y^2-k_0^2)=ik_0\kappa\left[1+4\pi^2\tilde{\sigma}_{xx}\tilde{\sigma}_{yy}\right]/2\pi$,
written in terms of the dimensionless conductivity components $\tilde{\sigma}_{ii}=\sigma_{ii}/c$, where $k_0=\omega/c$. The out-of-plane
component of the plasmon wave vector is $q_z=i\kappa$,  $\kappa(\omega)=\sqrt{q_x^2+q_y^2-k_0^2}$, where $\kappa^{-1}$ measures the penetration
depth of the evanescent mode into the neighbouring media,
which we assume to be vacuum for the sake of simplicity.
When the power absorption is small, such that $\text{Re}[\sigma_{ii}(\omega)]\approx 0$, 
the dispersion relation reduces to a real equation.
For a given frequency, this corresponds to a fourth-degree algebraic curve, whose gradient vector field defines the direction of energy propagation.
Fig. \ref{fig4} illustrates the plasmon modes in ${\bf q}$-space for the cases shown in Fig. \ref{fig2}, at frequencies lying inside or outside a hyperbolic region.
Given that $q_x,q_y\gg k_0$, the dispersion relation
becomes well described by the expression 
\begin{equation} \label{DR2}
\frac{q_x^2}{\tilde{\sigma}''_{yy}(\omega)}+\frac{q_y^2}{\tilde{\sigma}''_{xx}(\omega)}=
2\pi k_0q\left\{[4\pi^2\tilde{\sigma}''_{xx}(\omega)\tilde{\sigma}''_{yy}(\omega)]^{-1} - 1\right\},
\end{equation}
where $\tilde{\sigma}''_{ii}\equiv\text{Im}[\tilde{\sigma}_{ii}(\omega)]$. 

We can identify the characteristic properties associated to hyperbolic propagation\cite{Alu,TLow2016}.
Closed quasi-elliptic contours are obtained whenever 
$\text{Im}[\sigma_{xx}(\omega)]\text{Im}[\sigma_{yy}(\omega)]>0$, elongated along the smallest of $\sigma_{ii}$, 
or hyperbolic branches otherwise, with asymptotes having slopes $\tan\theta=\pm\sqrt{-\text{Im}\,\sigma_{xx}(\omega)/\text{Im\,}\sigma_{yy}(\omega)}$. The group velocity points mainly along the directions $\mp\sqrt{-\text{Im}\,\sigma_{yy}(\omega)/\text{Im\,}\sigma_{xx}(\omega)}$, normals to the asymptotes, and large wave vectors can be supported, indicating a high degree of directionality and localization. This contrasts 
with the purely anisotropic case (closed curves) where there
is propagation in all directions, although less confinement of the mode. 
%{\color{red}
%[Esto debe ser consistente con lo observado en la longitud de propagaci\'on $\text{Re}(q)/\text{Im}(q)$]}
The topology (closed vs. open isofrequency contours) and
the direction of the plasmon propagation can be manipulated
not only by changing the exciting frequency, but also
varying the position of the Fermi energy only
(Figs. \ref{fig4}(b) and (c)).

Note that using borophene 8-$Pmmn$ as reference, all
the parameters in the model (\ref{H1}) are fixed, with the exception of the magnitude of the mass $\Delta$. This is in contrast to the minimal model considered in Ref.~\cite{TLow2016} where six parameters are varied, having multilayer black phosphorus as the model candidate for anisotropic 2D intrinsic material supporting hyperbolic modes. In our case, the calculated conductivity response can not be effectively parametrized as in Ref.~\cite{TLow2016}. Moreover, we found that even for isotropic velocity, $v_x=v_y$, the tilting and the indirect nature of the gap along $k_y$-direction provide sufficient anisotropy to the system, while the introduction of mass leads to important modifications of the interband contribution as mentioned above. The joint effect of these two elements are enough to produce an interplay of the intraband and interband excitations capable of making the appearance of hiperbolicity of plasmon propagation possible, without any further material parameters.

\begin{figure}
    \centering
    \includegraphics[scale=0.32]{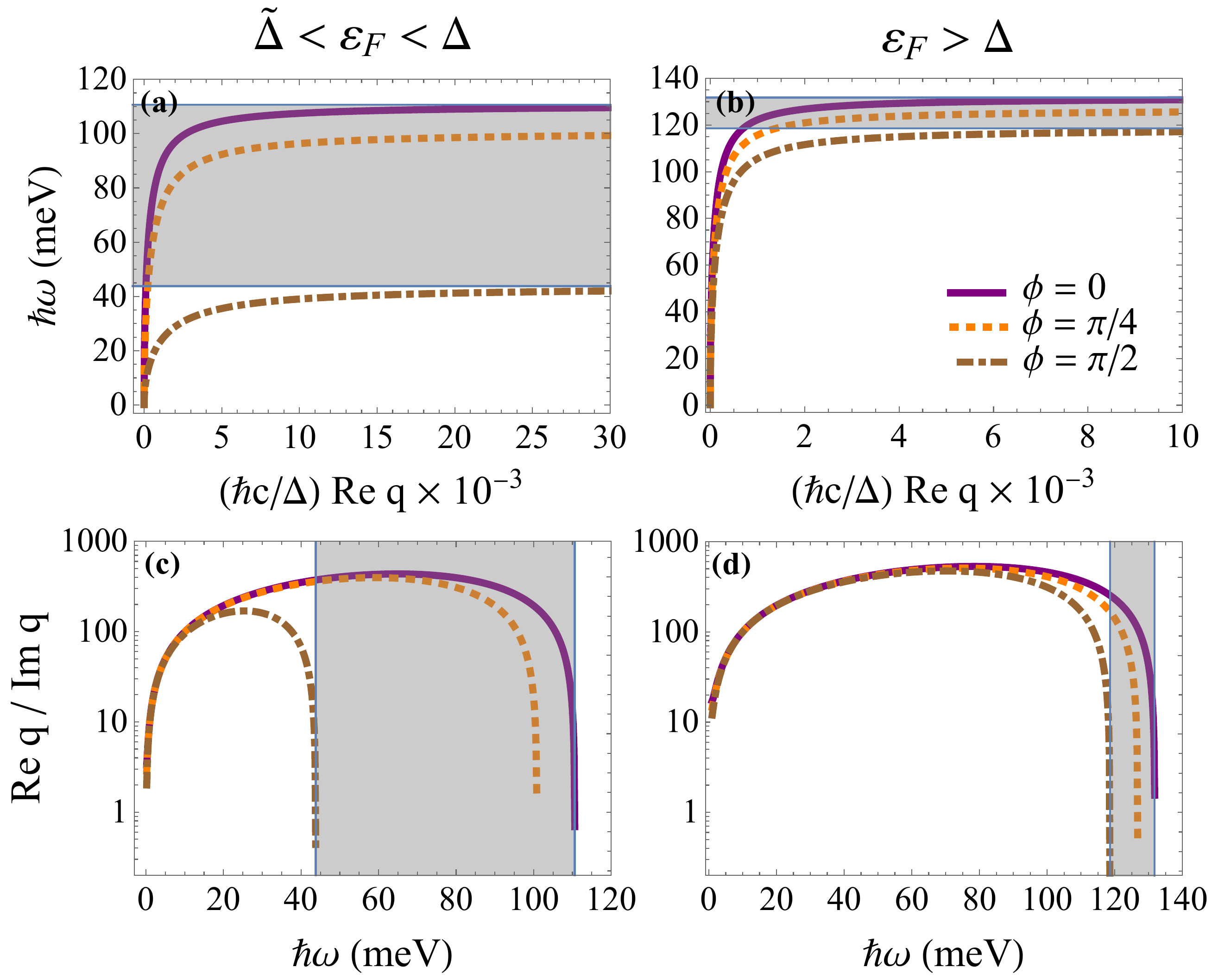}
    \caption{Frequency-momentum relation ((a), (b)) and inverse of losses  ((c), (d)) of the collectives modes for several propagation angles, when $\tilde{\Delta}<\varepsilon_F<\Delta$ and  $\varepsilon_F>\Delta$. Hyperbolic regions (in gray) correspond to those in Fig. \ref{fig2}.}
    \label{fig5}
\end{figure}

Figs. \ref{fig5}(a) and (b) display the surface plasmon dispersion for different angles of propagation $\phi$, at the same two positions of Fermi level as in Fig. \ref{fig2}. The increasing density of states of modes as the dispersion curve enters the zone of hyperbolicity reflects the increasing of spatial confinement.
Figs. \ref{fig5}(c) and (d) show the plasmon losses as it propagates within the surface along several directions $\phi$. For $\phi=\pi/2$, the dispersion
relation has a solution for frequencies below and up to the lower bound of the hyperbolic zone, as the existence of closed isofrequency contours illustrate (Figs. \ref{fig4}(a),(c)).
For smaller $\phi$, the solution enters a hyperbolic region eventually, and will present a frequency cutoff $\omega_c$ defined by the condition that the decreasing slope (as $\omega$ increases) of the asymptote  (see Fig.~\ref{fig2}) finally match the direction $\phi$; that is the equation $\theta(\omega_c)=\phi$. For $\phi=0$, $\omega_c$ is given by $\text{Im}[\sigma_{xx}(\omega_c)]=0$, which corresponds to the higher border of the hyperbolic zone. For $\phi=\pi/4$ the cutoff is red shifted because the frequency dominion available to reach the condition $\theta(\omega_c)=\phi$ is reduced. 
As $\phi\to\pi/2$, the solution cutoff approaches the frontier between elliptical and hyperbolical propagation,  
as $\text{Im}[\sigma_{yy}(\omega)]\to 0^-$.
On the other hand, at low frequencies the losses increase due to the intraband absorption. 
It is interesting to note that the propagative losses reach its minimum within the hyperbolic zone when $\tilde{\Delta}<\varepsilon_F<\Delta$, 
and within an elliptic zone when $\varepsilon_F>\Delta$. Thus, the system can support well defined, highly directional hyperbolic plasmons. In particular, at a given frequency, the position of the Fermi level allows to tune between purely anisotropic or hyperbolic propagation.
According to Hamiltonian (\ref{H1}), or from a previous knowledge of a nominal gap, one might think that the condition $\varepsilon_F<\Delta$ %means that the Fermi level lies within the gap, and therefore 
imply that hyperbolic propagation is impossible because the absence of intraband transitions. However, as Fig. \ref{fig5}(c) shows, the propagation of hyperbolic plasmon is still possible because the appearance of the indirect zone. This suggests that the Fermi energy dependence of the transitions illustrated in Fig.~\ref{fig5} is a clear indication of the presence of tilt and
mass in the system.

There is a proposal of generating contrasting gaps in 2D Dirac materials by magnetic impurities\cite{hill2011valley,Matis2016}. We have also performed calculations considering these valley-dependent gaps. In this situation, the additional breaking
of the time-reversal symmetry leads to a more complex
spectral behavior of the optical response, given the extra
possibilities for the position of the Fermi level and the associated spectrum of allowed excitations\cite{MRJ_1}. As a consequence, similar topological transitions in the type of
collective modes propagation, like those described above,
have been found, with an extended tunability.

In summary, we have shown how an anisotropic 2D system with broken inversion and particle-hole symmetries can support well-defined, highly collimated, hyperbolic surface electromagnetic modes in the THz range. In particular, simultaneous mass $and$ tilt lead to indirect gaps along the tilting direction,  an additional source of anisotropy that substantially modifies the interband contribution to the optical response. Consequently, regions of hyperbolicity arise for the exciting frequency, modulable by the appropriate location of the Fermi level or the magnitude of the tilting. This mechanism differs from those reported for natural hyperbolic 2D systems like black phosphorus or some transition metal dichalcogenides like WTe$_2$. The characteristic topological transitions between closed and open forms of plasmon propagation suggest an optical signature of the simultaneous occurrence of tilt and mass. Although motivated by systems like borophene 8-$Pmmn$ and 
$\alpha$-(BEDT-TTF)$_{2}$I$_{3}$, our study
should be of relevance to the optical properties of anisotropic atomically thin materials.

We acknowledge useful discussions with Catalina L\'opez Bastidas, V\'ictor Ibarra Sierra and Juan Carlos Sandoval. M.A.M and R.C.-B. thank 20va Convocatoria Interna (UABC) 400/1/64/20.

\bibliography{biblio.bib}
\end{document}